# Magnetic and electronic properties of nitrogen-doped lanthanum sesquioxide La$_2$O$_3$ as predicted from first principles

V.V. Bannikov • I.R. Shein • A.L. Ivanovskii

**Abstract** Using the *ab initio* FLAPW-GGA method we examine the electronic and magnetic properties of nitrogen-doped non-magnetic sesquioxide La$_2$O$_3$ emphasizing the role of doping sites in the occurrence of $d^0$-magnetism. We predict the magnetization of La$_2$O$_3$ induced by nitrogen impurity in both octahedral and tetrahedral sites of the oxygen sublattice. The most interesting results are that (i) the total magnetic moments (about 1 μ$_B$ per supercells) are independent of the doping site, whereas (ii) the electronic spectra of these systems differ drastically: La$_2$O$_3$:N with six-fold coordinated nitrogen behaves as a narrow-band-gap magnetic semiconductor, whereas with four-fold coordinated nitrogen is predicted to be a magnetic half-metal. This effect is explained taking into account the differences in N-2$p_z^{\downarrow\uparrow}$ *versus* N-2$p_{x,y}^{\downarrow\uparrow}$ orbital splitting for various doping sites. Thus, the type of the doping site is one of the essential factors for designing of new $d^0$-magnetic materials with promising properties.

V.V. Bannikov • I.R. Shein(✉)• A.L. Ivanovskii
Institute of Solid State Chemistry, Ural Branch of the Russian Academy of Sciences, 620990, Ekaterinburg, Russia
E-mail:  shein@ihim.uran.ru

## 1 Introduction

In recent years, the so-called $d^0$ magnets [1] became one of the most intensively investigated groups of materials owing to their promising applications in opto- and spintronics, which were considered both theoretically and experimentally, see reviews [2-4].

The magnetic properties ot these materials appear from spin polarization of *p* orbitals, and this effect is the strongest for 2*p* atoms (carbon, nitrogen or oxygen), for which the Hund energy is close to that of transition metal atoms. In turn, except some perfect crystals such as intrinsic $d^0$ magnets (highest oxides such as RbO$_2$, Rb$_4$O$_6$, Cs$_4$O$_6$ or several II–V and II–IV compounds such as BaN, SrN, BaC, CaP *etc.*, see [2-5]), three main types of occurrence of $d^0$ magnetism are known today: so-called impurity-induced, vacancy-induced and of mixed type: (impurity+vacancy)-induced magnetism [2-4,6,7].

Recent theoretical calculations predicted the appearance of $d^0$ magnetism in a rather broad family of semiconducting or insulating oxides and nitrides (such as BeO, CaO, MgO, ZnO, HfO$_2$, TiO$_2$, SnO$_2$, AlN, InN etc.), which is induced by various lattice defects (lattice vacancies or (and) by 2*p* dopants: B, C, N or O); these predictions are indeed confirmed in recent experiments discussed in [2-7]. We underline that the majority of these materials are based on high-symmetric crystals (mainly, cubic), where lattice defects occupy a unique type of atomic positions, and the variations of their magnetic properties were examined [2-7] as depending on (i) the type of lattice defects, (ii) their concentration and on (iii) ordering of these defects.

However, the family of $d^0$ magnets with interesting physical properties may be significantly enlarged, involving more complex wide-band-gap semiconductors and insulators such as metal sesquioxides $M_2O_3$ - by chemical substitutions (doping effects) or by introduction of lattice vacancies (non-stoichiometry effects) *in various non-equivalent atomic positions* of the matrix.

In this Letter, using the *ab initio* FLAPW-GGA method we examine the electronic and magnetic properties of nitrogen-doped non-magnetic sesquioxide $La_2O_3$, focusing on the role of doping sites in the occurrence of $d^0$-magnetism. The data of [8-10] reporting successful synthesis of lantanum oxynitrides are the experimental background for choosing the $La_2O_3$:N system in our analysis.

## 2 Models and computational details

The most stable lanthanum sesquioxide polymorph, α-$La_2O_3$, adopts a trigonal (space group *P*-3*m*1, Z = 1) structure, where two La atoms are placed in equivalent positions (1/3;2/3;*u*) and (2/3;1/3;1-*u*). The oxygen atoms are in two nonequivalent positions with six-fold (distorted octahedral, $O^o$, with $D_{3h}$ symmetry) and four-fold (distorted tetrahedral, $O^t$, with $C_{3v}$ symmetry) atomic coordination. The positions of oxygen atoms are: for one $O^o$ atom: (0;0;0), and for two $O^t$ atoms: (2/3;1/3;*v*) and (1/3;2/3;1-*v*), where *u* and *v* are so-called internal coordinates.

We simulate the nitrogen-doped α-$La_2O_3$ using a 2×2×1 supercell, where a nitrogen atom is introduced either into an octahedral (0;0;0) position (*o*-$La_2O_3$:N) or into a tetrahedral (2/3;1/3;*u*) position (*t*-$La_2O_3$:N), see Fig. 1. In this way the nominal compositions $La_2O_{2.75}N_{0.25}$ is simulated.

Our calculations were performed by means of the full-potential linearized method of augmented plane waves (FP-LAPW) with mixed basis APW+lo implemented in the WIEN2k suite of programs [11].

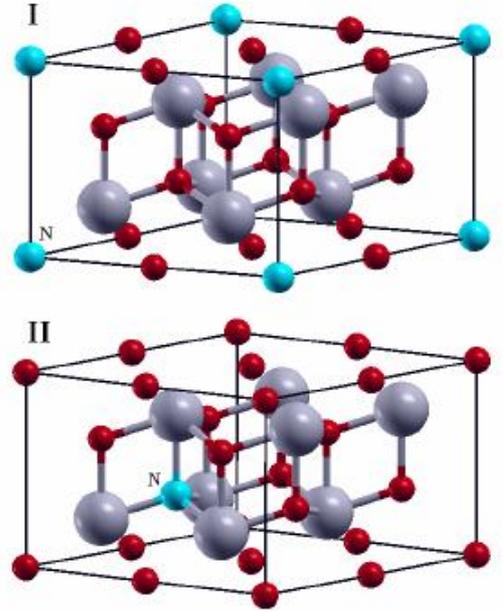

**Fig. 1** (*Color online*) The atomic structures used in our simulations of $La_2O_3$ with nitrogen impurity introduced into distorted octahedral (*o*-$La_2O_3$:N; I) and distorted tetrahedral sites of oxygen sublattice (*t*-$La_2O_3$:N; II).

The generalized gradient approximation (GGA) to exchange-correlation potential in Perdew-Burke-Ernzerhof form [12] was used. The calculations were performed taking plane-wave expansion with $R_{MT} \times K_{MAX}$ equal to 7 and 7×7×8 *k*-mesh (52 *k*-points in the irreducible part of Brillouin zone). Relativistic effects were taken into account within the scalar-relativistic approximation. The MT radii were chosen equal to 2.30 a.u. for La and 2.00 a.u. both for oxygen and nitrogen atoms. The self-consistent calculations were considered to have converged when the difference in the total energy of the crystal did not exceed 0.1 mRy as calculated at consecutive steps. The integration over the Brillouin zone was performed employing the modified tetrahedron method [13].

## 3 Results and discussion.

As the first step, the structure of pure α-$La_2O_3$ was optimized. The equilibrium lattice constants and internal coordinates were calculated to be $a$ = 3.936 Å, $c$ = 6.127 Å, $u$ = 0.240 and $v$ = 0.351, in agreement with experiments ($a_{exp}$ = 3.937 Å, $c_{exp}$ = 6.129 Å; JCPDS no. 05-0602) and earlier calculations [14,15]. Then, according our band structure calculations, the filled valence band of $La_2O_3$ is mainly of the O-2$p$ character, and the bottom of the empty conduction band is composed predominantly of La-5$d$,4$f$ states – in accordance with earlier data [14,15]. The obtained band gap is ~ 3.9 eV, which agrees with the results of other calculations [15], but (as well as for all other LDA/GGA calculations) is underestimated as compared with the experimental value 5.8 ÷ 6.2 eV [16,17].

Let us discuss the electronic properties of the N-doped lanthanum sesquioxide. Firstly, we found that for both systems: $o$-$La_2O_3$:N and $t$-$La_2O_3$:N the magnetic ground state becomes energetically favorable. In general, the appearance of $d^0$ magnetism in $La_2O_3$:N is similar to that in some other insulating oxides doped with 2$p$-impurities and originates from spontaneous spin-polarization of N-2$p$ orbitals, see [2-7,18,19].

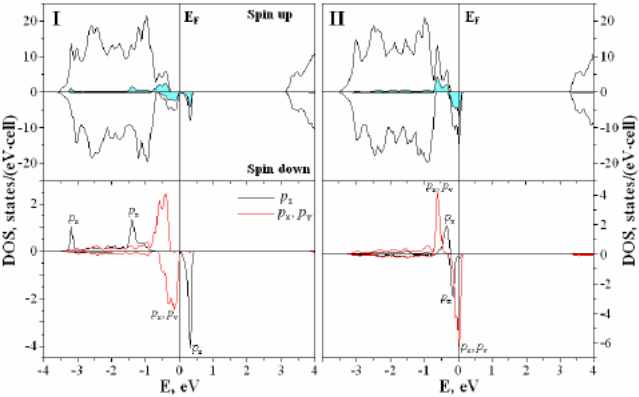

**Fig. 2** (*Color online*). Spin-resolved total (*upper panel*) and N $p_z^{\downarrow\uparrow}$, $p_{x,y}^{\downarrow\uparrow}$ partial densities of states (*bottom panel*) for $La_2O_3$ with nitrogen impurity introduced into distorted octahedral ($o$-$La_2O_3$:N; I) and distorted tetrahedral sites of oxygen sublattice ($t$-$La_2O_3$:N; II). On the upper panel, the N-2$p$ DOSs are also depicted (*shaded*).

Namely, from the total densities of states (DOSs) of $o$-$La_2O_3$:N and $t$-$La_2O_3$:N shown in Fig. 2, it is seen that substitutional doping with nitrogen in octahedral or tetrahedral oxygen sites results in: (i). appearance of new impurity N-2$p$ bands localized in the gap of the lanthanum sesquioxide; (ii). a shift of the Fermi level $E_F$, which is located in the region of these N-2$p$-like impurity bands; and (iii) splitting of these impurity bands in two spin-resolved N-2$p^{\downarrow\uparrow}$ bands.

Secondly, appreciable magnetic moments (MMs) appear in these systems on nitrogen atoms: 0.67 $\mu_B$ for $o$-$La_2O_3$:N and 0.57 $\mu_B$ for $t$-$La_2O_3$:N. Besides, small MMs 0.03-0.07 $\mu_B$ are induced on the oxygen atoms nearest to the impurity. The estimated total magnetic moments (per supercells) are about 1 $\mu_B$. Thus, total magnetization of the $La_2O_3$:N system remains almost independent of the doping site.

The most interesting result is that depending on the doping site, the electronic spectra of $o$-$La_2O_3$:N *versus* $t$-$La_2O_3$:N are very different. Indeed, upon substitution of a nitrogen atom for a six-fold oxygen $O^o$ by, the N-2$p^{\uparrow}$ band is located below the Fermi level and is filled, whereas the N-2$p^{\downarrow}$ band splits in two subbands, one of which lies below $E_F$ and is also filled, and the second - is located above of $E_F$ and remains empty, Fig. 2. As a result, $o$-$La_2O_3$:N will behave as a ***magnetic semi-conductor*** with zero density of carriers at the Fermi level for both spin projections $N^{\downarrow\uparrow}(E_F)$ = 0. The calculated band structure of $o$-$La_2O_3$:N is shown in Fig. 3. We find that this system is a narrow-band semiconductor with direct transition at the A point and with the GGA gap at about 0.05 eV.

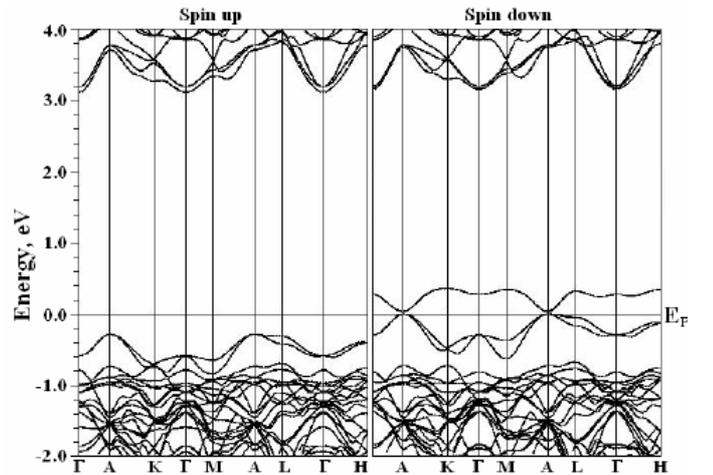

**Fig. 3** Spin-resolved electronic bands of $La_2O_3$ with nitrogen impurity introduced into distorted octahedral sites

of oxygen sublattice (o-$La_2O_3$:N).

On the contrary, for t-$La_2O_3$:N the N-$2p^\uparrow$ band is also located below the Fermi level $E_F$ and is filled, whereas the Fermi level crosses the N-$2p^\downarrow$ band, Fig. 2. As a result, t-$La_2O_3$:N is characterized by nonzero density of carriers at the Fermi level for one spin projection ($N^\downarrow(E_F) > 0$), but has a band gap for the reverse spin projection ($N^\uparrow(E_F) = 0$). As is known, such types of spectra are typical of the so-called *magnetic half-metals* (MHM), for which spin density polarization at the Fermi level is $P = \{N^\downarrow(E_F) - N^\uparrow(E_F)\}/\{N^\downarrow(E_F) + N^\uparrow(E_F)\} = 1$ and the conduction occurs along preferred spin channels, determining nontrivial spin-dependent transport properties of MHM materials, see [2-4].

The above two main results for $La_2O_3$:N, *i.e.* the invariance of the magnetic moments simultaneously with drastic differences in electronic spectra due to substitutional doping at octahedral or tetrahedral oxygen sites, may be explained taking into account the transformations of N-$2p$ bands as depends on the type of La surrounding. It is well known that the crystal field of cubic symmetry does not split the $^2P$ term ($p_x$, $p_y$, and $p_z$ components form the irreducible representation $T_{1u}$ of point group $O_h$ or $T_2$ of the tetrahedral group), meanwhile its trigonal distortions (of $C_{3v}$ and $D_{3h}$ symmetry) lead to splitting of this term into non-degenerate $p_z$ and doubly degenerate $p_{x,y}$ components. The decomposition of the N-$2p$ band into $p_z$ and $p_{x,y}$ components is shown in Fig. 2. We see that owing to different coordination numbers and N-La distances, the splitting of $p_z^{\downarrow\uparrow}$ and $p_{x,y}^{\downarrow\uparrow}$ components in crystal field for o-$La_2O_3$:N and t-$La_2O_3$:N is quite different. For o-$La_2O_3$:N, the $p_{x,y}^{\downarrow\uparrow}$ states show relatively small spin splitting (at about 0.4 eV) and are fully occupied. On the contrary, the $p_z^{\downarrow\uparrow}$ states are divided into two well separated spin-resolved $p_z^\downarrow$ and $p_z^\uparrow$ bands, responsible for the formation of atomic MM, where the $p_z^\uparrow$ states are below, and the $p_z^\downarrow$ states above $E_F$.

A quite different situation was found for t-$La_2O_3$:N, Fig. 2. Here, when N is placed in the $O^t$ site, the $p_z^{\downarrow\uparrow}$, $p_{x,y}^{\downarrow\uparrow}$ orbital splitting becomes of the opposite type as compared with o-$La_2O_3$:N, *i.e.* the $p_z^{\downarrow\uparrow}$ states are below $E_F$ and are fully occupied, whereas the highest splitting is demonstrated by $p_{x,y}^\downarrow$ (*versus* $p_{x,y}^\uparrow$) orbitals. In particular, the Fermi level bisects the minority $p_{x,y}^\downarrow$ band, resulting in half-metallicity of t-$La_2O_3$:N.

Finally, let us note that the total energy difference between the considered o-$La_2O_3$:N and t-$La_2O_3$:N was estimated to be about 0.1 eV per furmula unit. Thus, for the doped sesquioxide $La_2O_3$:N the magnetic semi-conducting state induced by nitrogen impurity introduced into octahedral sites will be expected as the preferable one.

## 4 Conclusions

In summary, the electronic and magnetic properties of the nitrogen-doped sesquioxide $La_2O_3$ have been studied by the FLAPW-GGA approach, with focus on the role of doping sites in the occurrence of $d^0$-magnetism.

We found the magnetization of $La_2O_3$ induced by a nitrogen impurity introduced either into octahedral or tetrahedral sites of the oxygen sublattice. In both cases independently of the doping site the total magnetic moments coincide (at about $1\mu_B$, per supercells), whereas the electronic spectra of these systems differ drastically. Namely, the $La_2O_3$:N system with six-fold coordinated nitrogen behaves as a magnetic semi-conductor, whereas with four-fold coordinated nitrogen - as a magnetic half-metal. This effect is explained taking into account the differences in the splitting character of $p_z^{\downarrow\uparrow}$ *versus* $p_{x,y}^{\downarrow\uparrow}$ orbitals for various doping sites.

Finally, we believe that the type of doping site belongs to the most essential factors (together with the type of lattice defects, their concentration and ordering type known earlier) for designing of new magnetic materials with promising properties.

**Acknowledgments** Financial support from the RFBR (Grant 10-03-96004-Ural) is gratefully acknowledged.

# 5 References

1. Coey, J.M.D.: Solid State Sci. **7,** 660 (2005)
2. Katayama-Yoshida, H., Sato, K., Fukushima, T., Toyoda, M., Kizaki, H., Dinh, V.A., Dederichs, P.H.: phys. stat. sol. (a) **204**, 15 (2007).
3. Ivanovskii, A.L.: Physics-Uspekhi **50**, 1031 (2007)
4. Volnianska, O., Boguslawski, P.: J. Phys.: Condens. Matter **22**, 073202 (2010)
5. Kováčik, R., Ederer, C.: Phys. Rev. B **80**, 140411R (2009)
6. Shein, I.R., Gorbuniva, M.A., Makurin, Yu.N., Kiiko, V.S., Ivanovskii, A.L.: Intern. J. Modern Phys. **22**, 4987 (2008).
7. Ivanovskii, A.L., Shein, I.R., Makurin, Yu.N., Kiiko, V.S., Gorbunova, M.A.: Inorgan. Mater. **45**, 223 (2009)
8. Brown, R.C., Clark, N.J.: J. Inorgan. Nuclear Chem. **36**, 2287 (1974)
9. J. Rooke, J., W. Weller, W. in: Solid State Chemistry V (Book Ser.: Solid State Phenomena), **90-91**, 417 (2003).
10. Ebbinghaus, S.E., Abicht, H.P., Dronskowski, R., Muller, T., Reller, A., Weidenkaff, A.: Progr. Solid State Chem. **37**, 173 (2009)
11. Blaha, K. Schwarz, G. Madsen et al. WIEN2k, An Augmented Plane Wave Plus Local Orbitals Program for Calculating Crystal Properties. Vienna University of Technology, Vienna, 2001.
12. Perdew, J.P., Burke, K., Ernzerhof, M.: Phys. Rev. Lett. **77**, 3865 (1996)
13. Blochl, P.E., Jepsen, O., Anderson, O.K.: Phys. Rev. B **49**, 16223 (1994).
14. Vali, R., Hosseini, S.M.: Comput. Mater. Sci. **31**, 125 (2004).
15. Mikami, M., Nakamura, S.: J. Alloys Comp. **408–412**, 687 (2006).
16. Robertson, J., Xiong, K., Clark, S.J.: Thin Solid Films, **496**, 1 (2006).
17. Xiong, K., Robertson, J.: Microelectronic Engin. **86**, 1672 (2009).
18. Bannikov, V.V., Shein, I.R., Kozhevnikov, V.L., Ivanovskii, A.L.: J. Magnet. Magn. Mater. **320**, 936 (2008).
19. Shein, I.R., Enyashin, A.N., Ivanovskii, A.L.: Phys. Rev. B **75**, 245404 (2007).